\documentclass[12pt]{article}
\usepackage{amsmath,a4}
\usepackage{epsfig}

\begin{document}
\begin{titlepage}
\begin{flushright}
IJS-TP-99/08\\ 
\end{flushright}

\begin{center}
{\Large \bf  The role of $K^*_0(1430)$ in 
$D \to P K$ and $\tau \to K P \nu_\tau$ decays 
  }\\
\vspace{1cm}

{\large \bf S. Fajfer$^{a,b}$ and J. Zupan$^{a}$ }\\

{\it a) J. Stefan Institute, Jamova 39, P. O. Box 3000, 1001 Ljubljana, 
Slovenia}
\vspace{.5cm}\\

{\it b) 
Department of Physics, University of Ljubljana, Jadranska 19, 1000 Ljubljana,
Slovenia}
\vspace{.5cm}

\end{center}

\vspace{0.25cm}

\centerline{\large \bf ABSTRACT}

\vspace{0.2cm}
We consider the scalar form factor in the 
weak current matrix element \break 
$<P K| j_{\mu} |0> $, $P = \pi, \eta, \eta'$.
It obtains the contributions from 
 the scalar meson resonance $K^*_0(1430)$ and    
from the 
scalar projection of the vector meson $K^*(892)$ resonance. 
We analyze  decay  amplitudes 
of the Cabibbo suppressed decays $D\to K P$, $P = \pi, \eta, \eta'$ using the  
factorization approach. The form factors of the relevant matrix elements 
are described by assuming the dominance of  nearby resonances. 
The annihilation 
contribution in  these decays  
arises from the  matrix element 
$<P K| j_{\mu} |0> $.  
 All the required parameters are experimentally known except the scalar meson 
$K^*_0(1430)$ decay constant.   We fit the decay amplitudes and we 
find that final state interaction  improves the agreement with the 
experimental data. 
Then  we extract  bounds on  scalar form factor parameters 
and  compare them  with  the experimental data 
obtained in the analyses of  $K \to \pi e \nu_e$ and $K \to \pi \mu \nu_\mu$. 
The same scalar form factor is present in the $\tau \to K P \nu_\tau$ 
decay, with $P = \pi, \eta, \eta'$. 
Using the obtained bounds  we investigate the significance of the scalar 
meson form factor  in  the $\tau \to K P \nu_\tau$, $P = \pi, \eta, \eta'$ 
decay rates and   
spectra.  We find that  the $K^*_0(1430)$ scalar meson dominates in the 
 $\tau \to K \eta'\nu_\tau$ decay spectrum.\\
\vskip1cm
PACS number(s): 13.25.Ft, 13.35.Dx, 12.15.-y
\end{titlepage}

\def\diag{\operatorname{diag}}
\def\Tr{\operatorname{Tr}}
\section{Introduction}
The matrix element of the weak current in the $K_{l3}$ decays 
is usually described by the use of vector and scalar form factors  
\cite{PDL-98} 
 \begin{equation}
\begin{split}
 <\pi^0(p)| \bar s \gamma_{\mu} (1 - \gamma_5) u |K^+(p')>= 
&F_+ (q^2) \left(p'_\mu + p_\mu 
-\frac{m_K^2-m_\pi^2}{q^2}(p'_\mu-p_\mu)\right) + \\
&+F_0 (q^2) \frac{m_K^2-m_\pi^2}{q^2}(p'_\mu - p_\mu) . 
\end{split}\label{I-1}
\end{equation}
Analyses of $K_{l3}$ data assume a linear dependence on $q^2 = (p'-p)^2$ :
\begin{equation}
F_{+,0}^{K,\pi}(q^2)  = F_{+,0}^{K,\pi}(0) (1 + \lambda_{+,0}\frac{ 
q^2}{m_\pi^2}),
\label{e1}
\end{equation}
with the experimental fit for $\lambda_{+}$ compatible with $\lambda_+ \sim 
0.03$ both from $K_{e3}$ and  
$K_{\mu3}$ decays \cite{PDL-98}. 
The data for $\lambda_{0}$ are rather inconclusive as summarized in 
\cite{PDL-98},
with $\lambda_{0} = 0.006\pm 0.007$ obtained from $K_{\mu 3}^+$ decays, while 
$K_{\mu 3}^0$ decay experiments prefer $\lambda_0=0.025\pm0.006$. 
The calculation done within chiral perturbation theory  (CHPT) 
found  $\lambda_{0} = 0.017\pm 0.004$  \cite{Bijnens-92}.

The same matrix element $<PK | j_{\mu}(0) | 0>$, $P = \pi, \eta, \eta'$ might 
appear 
as  one of the contributions in some of the D meson nonleptonic decay 
amplitudes 
\cite{Bauer-87,Buccella-95}, 
as well as in the $\tau \to K P \nu_\tau$ decays.

A satisfactory explanation of the mechanism of 
exclusive nonleptonic weak D meson  decays has not been found yet. 
The method of heavy quark expansion cannot be  used  
successfully on charm sector due to too small 
$c$ quark mass. The simple widely used 
factorization ansatz for the matrix elements of the amplitude does 
 not explain properly the Cabibbo allowed $D^0$ decays 
\cite{Bauer-87}-
\cite{Kamal-96a}. 
There are some attempts to approach the nonfactorized contribution 
\cite{Kamal-96}-
\cite{Smith-98}. However,  
the factorization technique \cite{Bauer-87,Buccella-95,Verma-95,Cheng-99} 
is mostly used  to treat  the $D$ meson decay amplitudes. 
 Using this approach  
 the amplitude of the D meson nonleptonic decay 
 is divided into well known 
{\it  spectator } and 
{\it annihilation} contribution \cite{Bauer-87,Buccella-95,Verma-95}. 
Contrary to the quite well understood spectator contribution the annihilation 
contribution is rather poorly known. 
 
In \cite{Buccella-95} an attempt has been made to treat the D meson 
nonleptonic 
decays assuming that the final state interaction (FSI) is dominated by nearby 
resonances. In this approach 
the annihilating contribution was fitted directly and the authors of 
\cite{Buccella-95} 
 noticed that this  contribution is larger than one would get by assuming the 
  $K^*_0(1430)$ dominance of the scalar form factor 
  with parameters fixed from $K\to \pi l \nu_l$ decays \cite{Bauer-87}. 
  
Motivated by this attempt we reinvestigate the $D^0 \to \bar K^0 P$, 
$P = \pi^0, \eta, \eta'$ and $D^0 \to K^-  \pi^+$ 
decays in which the annihilation contribution is proportional to the 
scalar form factor of the matrix element $<P K| j_{\mu} |0>$. 
We assume the  $K^*_0(1430)$  meson 
dominance of the scalar form factor and we include the scalar projection of
 the vector meson $ K^*(892)$  resonances following the idea of 
\cite{Finkemeier-96}.

In our numerical calculation we use all existing data on $\tau \to K^* (892) 
\nu_\tau$, $K^*(892)\to K\pi$, $K^*_0(1430) \to K \pi$. 
Unfortunately,  the present experimental data on $K_{l3}$ decay \cite{PDL-98} 
cannot give reliable bounds on the scalar meson decay constant. Therefore, 
we make fit using the 
decay rates of the $D^0 \to \bar K^0 P$, $P = \pi^0, \eta, \eta'$ and $D^0 \to 
K^-  \pi^+$ data.  
 We find better agreement with the experimental data if additional 
final state interaction  is taken into account. In treatment of FSI 
we follow  the  idea of 
 \cite{Buccella-95} 
and then extract bounds on the  $K^*_0(1430)$ decay constant. 
Making the low-energy expansion of our scalar form factor, we then compare 
our result
with the  $K_{l3}$ data and the chiral perturbation theory result. 

The scalar form factor in the  
$\tau \to K \pi  \nu_\tau$ decay has motivated a number of studies 
\cite{Finkemeier-96,Truong-95,Lichard-97}. In \cite{Finkemeier-96} 
 the scalar  form factor was approached by exchange of 
resonance $K^*_0(1430)$ noticing the presence of  the scalar projection of 
off-shell vector resonances $K^*(892)$.   
Using bounds on the scalar form factor in the \break 
$<P K| j_{\mu} |0>$ matrix 
element, 
determined from the study of the  D nonleptonic decays, 
we calculate the $\tau \to K P \nu_\tau$, $P = \pi, \eta, \eta'$, 
 decay spectra and widths. We notice that the 
scalar meson contribution is  rather small in 
$\tau \to K P l \nu_l$, $P = \pi, \eta$, while it gives the 
dominant contribution in the $\tau \to K \eta'  \nu_\tau$ decay rate. 

In Sec. 2 we reinvestigate the annihilation contribution
of the $D^0 \to KP$, $P=\pi, \eta, \eta'$  decays. 
In Sec. 3 we analyze the effects of the  
final state interaction in  these decay amplitudes. Sec. 4 is devoted to
 $\tau$ 
decays. 
Short summary of our 
results is given in Sec. 5.

\section{Annihilation contribution in the D meson nonleptonic 
decays}\label{Decay-amplitudes}

We briefly review the use of factorization approximation in the weak 
nonleptonic  D meson decays
 \cite{Bauer-87,Buccella-95,Verma-95,Cheng-99,Bigi-95,Kamal-96a,Buccella-90}.
The effective weak Hamiltonian for nonleptonic decays of charmed particles is 
given by
\begin{equation}
H_{\text{eff}}^w=\frac{G_f}{\sqrt{2}}( \sum_{q_1q_2} V_{cq_1}^* V_{uq_2} 
[a_1(\bar{q}_1
c)^{\mu}(\bar{u}q_2)_\mu +a_2 (\bar{q}_1q_2)_{\mu}(\bar{u}c)^\mu]+h.c.) ,
\label{art-2}
\end{equation}
where $q_{1,2}$ stand for $s$ or $d$ quark operators,
 while currents $\bar{q}_1 \gamma_\mu (1-\gamma_5)q_2$ are abbreviated as 
 $(\bar{q}_1q_2)_\mu$ and $V_{qq'}$ are the CKM matrix elements. 
 In the factorization approximation approach the invariant amplitude is 
 written in terms of current 
matrix elements between vacuum and meson states
\begin{equation}
\begin{split}
M_{P_1P_2,D}^W  =\hskip15mm&\\
\frac {G_f}{\sqrt 2} \sum_{q_1q_2}V_{cq_1}^*V_{uq_2} [&a_1\Big(\langle P_2| 
(\bar{u}q_2)_\mu | 0\rangle \langle 
P_1| (\bar{q_1} c)^\mu |D\rangle +\langle P_1| (\bar{u}q_2)_\mu | 
0\rangle\langle 
P_2| (\bar{q_1} c)^\mu |D\rangle+ \\
&\langle P_1P_2| (\bar{u}q_2)_\mu | 0\rangle\langle 0| (\bar{q_1} c)^\mu 
|D\rangle\Big)+ a_2\Big(\langle P_2| (\bar{q_1}q_2)_\mu | 0\rangle\langle P_1| 
(\bar{u} c)^\mu |D\rangle +\\
&\langle P_1| (\bar{q_1}q_2)_\mu | 0\rangle\langle P_2| (\bar{u} c)^\mu 
|D\rangle 
+\langle P_1P_2| (\bar{q_1}q_2)_\mu | 0\rangle\langle 0| (\bar{u} c)^\mu 
|D\rangle\Big)]. \label{fakt-2}
\end{split}
\end{equation} 
Here $a_1=1.26\pm0.04$ and $a_2=-0.51\pm0.05$ are effective Wilson
 coefficients \cite{Bauer-87}.

 Through Lorentz covariance we can write the most general expressions 
for matrix
elements 
\begin{equation}
\begin{split}
\left\langle P_1 (p_1)\right| V^\mu \left| P_2 (p_2) \right\rangle = 
&F_+^{(P_2 \rightarrow P_1 )} (q^2) \left(p_1^\mu+p_2^\mu 
-\frac{m_2^2-m_1^2}{q^2}q^\mu\right) + \\
&+F_0^{(P_2  \rightarrow P_1 )} (q^2) \frac{m_2^2-m_1^2}{q^2}q^\mu , 
\label{razp-3}
\end{split}
\end{equation}
\begin{equation}
\left\langle 0 \right| A^\mu \left| P(p)\right\rangle =  i\> f_P p^\mu  
\label{mat-2},
\end{equation}
\begin{equation}
\left\langle 0 \right| V^\mu \left| S(p)  \right\rangle = f_S p^\mu , 
\label{mat-3}
\end{equation}
\begin{equation}
\left\langle 0 \right| V^\mu \left| R(p,\lambda)  \right\rangle =i F_{R} 
\epsilon^\mu(p,\lambda) ,
\label{art-3}
\end{equation}
where $V^\mu$ and $A^\mu$ are vector and axial currents,
 $g^\mu=p_1^\mu-p_2^\mu$, while $P$, $S$, $R$ denote pseudoscalar, scalar 
and vector mesons respectively.
Defining $\langle P_1P_2|(\bar{q}'_1q_2)_\mu|0\rangle= J_\mu^{12} $ 
one obtains from \eqref{razp-3} \cite{Finkemeier-96,Choi-98} 
\begin{equation}
\begin{split}
J^{12}{}^\mu = 
&F_+^{(P_2 \rightarrow P_1 )} (q^2) \left(p_1^\mu-p_2^\mu 
-\frac{m_1^2-m_2^2}{q^2}(p_1^\mu+p_2^\mu)\right) + \\
&+F_0^{(P_2  \rightarrow P_1 )} (q^2) \frac{m_1^2-m_2^2}{q^2}(p_1^\mu
+p_2^\mu) .
\end{split}\label{art-1}
\end{equation}
To evaluate form factors $ F_{0,+}^{(P_2 \rightarrow P_1 )}(q^2)$ in 
\eqref{art-1} one usually uses single pole approximation 
\cite{Bauer-87,Lichard-97,Buccella-90,Ayala-87} 
from which one concludes that  annihilation term contribution to decay width 
is negligible. It was noticed in \cite{Buccella-95} that this is  not the case 
for 
two particle nonleptonic 
$D$ meson decays, where final state has  $S\ne 0$. 
This contribution was included in \cite{Buccella-95} 
by expressing the annihilation current 
densities $\partial^\mu J_\mu^{12}$ in terms of unknown variables multiplied 
by the $SU(3)$ Clebsch-Gordan coefficients. The magnitude of the 
coefficients turns out to be considerably larger than what one 
would obtain assuming the single pole approximation with parameters fixed from
 $K\to \pi l \nu_l$ decays \cite{Bauer-87}.
Here we reinvestigate the size 
of the annihilating contribution 
by assuming the meson dominance of the $\partial^\mu J_\mu^{12}$ matrix 
element.

In the analysis \cite{Finkemeier-96} of the $\tau \to K \pi \nu_{\tau}$ decay 
the scalar form factor received the contribution from the 
scalar projection of the
off-shell vector resonances and from  the scalar resonance $K_0^*(1430)$. 
In our  present approach we include this 
contribution of  
$K^*(892)$ and the contribution of the $K_0^*(1430)$ scalar meson.

In order to use the existing experimental data, we use following properties of 
the resonances. The $SU(3)$ symmetric effective strong Hamiltonian  
describing the decays of $K^*(892)\to K P$, and 
$K_0^*(1430) \to K P$, with $P = \pi, \eta, \eta'$ 
is given by
\begin{equation}
H_{\text{eff}}^S=ig_V \Tr ([\partial_\mu \Pi, \Pi] V^\mu) +g_S \Tr(\Pi \Pi S),
\label{art-5}
\end{equation}
where $\Pi$, $V$, $S$ are $3\times 3$ matrices
\begin{equation}
\Pi=
\begin{pmatrix}
\frac{\pi^0}{\sqrt 2}+\frac{\eta_8}{\sqrt 6} +\frac{\eta_0}{\sqrt 3} & \pi^+ & 
K^+\\
\pi^- &-\frac{\pi^0}{\sqrt 2}+\frac{\eta_8}{\sqrt 6} +\frac{\eta_0}{\sqrt 3}& 
K^0\\
K^- & \bar{K}^0 & -\frac{2}{\sqrt 6}\eta_8 +\frac{1}{\sqrt 3}\eta_0
\end{pmatrix}, \label{mocp-19}
\end{equation} 
while $S$ and $V$ have corresponding scalar or vector mesons as matrix 
elements.
The $SU(3)$ flavor symmetry breaking is taken into account through  
the physical masses and decay widths. 
The coupling constants $g_S$ and $g_V$ are some yet unknown coefficients 
that describe the strength of strong interaction.
In order to account as much as possible for the $SU(3)$ symmetry breaking 
effects we calculate relevant parameters  directly from the corresponding 
decays $K^*(892)$, $K^*_0(1430)$ and $\tau \to K^* \nu_\tau$, emphasizing this
 through the change of notation $g_V \to g_V(K^*)$, 
 $g_S \to g_S(K_0^*)$. In our analysis we did not 
include the  vector meson $K^*(1410)$ which presence was mentioned in 
\cite{Finkemeier-96} due to inducement of the large uncertainty in the vector 
form factor of the $\partial^\mu J_\mu^{12}$. 
Following \cite{Finkemeier-96}  we obtain  the current \eqref{art-1} 
for the intermediate $K^*(892)$ and $K_0^*(1430)$ meson states  

\begin{equation}
\begin{split}
J_\mu^{12}=&g_V(K^*) 2 a_{12K^*} F_{K^*} \frac{g_{\mu\nu}-
\frac{q_\mu q_\nu}{m_{K^*}^2}}{q^2-m_{K^*}^2 
+i\sqrt{q^2}\Gamma_{K^*}}(p_1^\nu-p_2^\nu)+\\
&f_{K_0^*}g_S(K_0^*) \frac{c_{12K_0^*}q^\mu}{q^2-m_{K_0^*}^2 
+i\sqrt{q^2}\Gamma_{K_0^*}},
\end{split}
\label{art-4}
\end{equation}
where $q=p_1+p_2$, while
\begin{align}
2a_{12K^*}(p_1-p_2)_\mu \epsilon^\mu &=\langle P_1P_2| \Tr([\partial_\mu\Pi,
\Pi]V^\mu)|K^*\rangle,\\
c_{12K_0^*}&=\langle P_1P_2|\Tr(\Pi\Pi S))|K_0^*\rangle.
\end{align}
The unknown coupling constant 
$F_{K^*}$ can be determined from the measured decay rate 
$\tau \to K^{*-}(892)\nu_{\tau}$, while $f_{K_0^*}$ is left as a free
 parameter.

The decay widths $\Gamma_X(q^2)$ in \eqref{art-4} are taken to be energy 
dependent \cite{Decker-93}
\begin{align}
\Gamma_X(q^2)&=\Gamma_X 
\frac{m_X^2}{q^2}\Big(\frac{p(q^2)}{p(m_X^2)}\Big)^{2n+1},\\
p(q^2)&=\frac{1}{2\sqrt{q^2}}\sqrt{[q^2-(m_{P_1}+m_{P_2})^2][q^2-(m_{P_1}-
m_{P_2
})^2]
},
\end{align}
where $n=1$ for the $K^*(892)$ (p-wave phase space) and $n=0$ for the $K_0^*$ 
(s-wave phase space)

The following expressions for form 
 factors are obtained by matching \eqref{art-4} with \eqref{art-1} 
\begin{align}
F_+^{(P_2\to P_1)}(q^2)=& \frac{g_V(K^*) 2 a_{12K^*} F_{K^*}} {q^2-m_{K^*}^2 
+i\sqrt{q^2}\Gamma_{K^*}},\\
F_0^{(P_2\to P_1)}(q^2)=& \frac{g_V(K^*) 2 a_{12K^*} F_{K^*}(1-q^2/m_{K^*}^2)}
 {(q^2-m_{K^*}^2 
+i\sqrt{q^2}\Gamma_{K^*})}+\frac{q^2}{(m_{P_1}^2-m_{P_2}^2)}\frac{f_{K_0^*}
g_S(K_0^*) 
c_{12K_0^*}}{(q^2-m_{K_0^*}^2 +i\sqrt{q^2}\Gamma_{K_0^*})}.
\label{art-7}
\end{align}

Note that in \eqref{art-4} as suggested by \cite{Finkemeier-96} we used a
 $K^*(892)$ propagator proportional to $(g_{\mu\nu}-q_\mu q_\nu/m_{K^*}^2)$ 
and not, as sometimes suggested, proportional to $(g_{\mu\nu}-
q_\mu q_\nu/q^2)$. The latter form would have led to a vanishing contribution 
of the vector resonance to the scalar form factor. However, this choice of 
propagator would fail to fulfill the condition $F_+(0)=F_0(0)$ and would thus 
lead to an unphysical divergence in current \eqref{art-1} as $q^2\to 0$. The 
contribution of vector meson to the scalar form factor on the other hand does 
vanish on the $K^*(892)$ mass-shell when $q^2=m_{K^*}^2$, this being in 
accordance with the expectations from dispersion relations \cite{Marshak-69}.

The  $K^*(892)$ meson  decays 
almost exclusively into $K \pi$ final state \cite{PDL-98}. 
Therefore we find using    \eqref{art-5} 
\begin{equation}
\Gamma(K^{*}(892)\to K\pi)=\frac{g^2_V(K^*) p_{K\pi}^3}{4\pi m_{K^*}^2}
\end{equation}
with rest frame momentum $p_{K\pi}$ of outgoing $K$ or $\pi$ meson
\begin{equation}
p_{K\pi}^2=\frac{m_{K^*}^2}{4}[(1-\frac{(m_K+m_\pi)^2}{m_{K^*}^2})
(1-\frac{(m_K-m_\pi)^2}{m_{K^*}^2})].\label{art-27}
\end{equation}
The value $\Gamma_{K^*}=50.8 \>{\rm MeV}$ \cite{PDL-98} yields 
 $g_V(K^*)=4.59$.

The value of $F_{K^*}$ is extracted from 
$\tau \to K^*(892) \nu_\tau$ decay width 
$\Gamma_{\tau \to K^* \nu_\tau}=2.905\cdot10^{-14} \>{\rm GeV}$ \cite{PDL-98}. 
Using 
\begin{equation}
\Gamma_{\tau \to K^*(892) \nu_\tau}=\frac{1}{4\pi m_\tau}G_f^2 F_{K^*}^2
(2+\frac{m_{\tau}^2}{m_{K^*}^2}) p^2 
\sin^2\Theta_C
\label{art-6}
\end{equation}
with  
$p={(m_\tau^2 -m_{K^*}^2)}/{(2m_\tau})$  
 we find $F_{K^*}=0.195 \>{\rm GeV}^2$.

In order to fix the $g_S(K_0^*)$ 
parameter we use
the total  decay width 
$\Gamma(K_0^*(1430))=287\pm23 \>{\rm MeV}$ and  branching ratio 
$B(K_0^*(1430) \rightarrow K \pi)=93 \pm 10 \% $ \cite{PDL-98}.  
Assuming that $K_0^*(1430)$  dominantly decays into 
$K P$, with $P = \pi, \eta$, instead of using nonet symmetry argument 
we allow in (\ref{art-5}) that scalar and psuedoscalar meson singlets have 
different couplings than octet states \cite{Buccella-95}. 
In order to implement the $\eta -\eta'$ mixing, we shall use the two-mixing 
angle formalism of the decay constants proposed by \cite{Leutwyler-98}, extended to the mixing of the $\eta$, $\eta'$ states  through $q^2$ dependent mixing angle in \cite{Escribano-99}. Thus one  has 
$|\eta> =\cos \theta_\eta |\eta_8> - \sin\theta_\eta|\eta_0>$ and 
$|\eta'> = \sin \theta_{\eta'} |\eta_8> +\cos \theta_{\eta'}|\eta_0>$.  The mixing angles 
were  found to be 
$\theta_{\eta}=-6.5^\circ\pm 2.5^\circ$ and $\theta_{\eta'}= -23.1^\circ\pm 3^\circ$ \cite{Escribano-99}.

The effective Hamiltonian describing the deviation from the nonet symmetry is 
given by
\begin{equation}
 {\cal H}_{\text{eff}}^S=g_{888} \Tr \big(\Pi_8^\dagger \Pi_8^\dagger
 {\rm S}_8\big) + 
\frac{1}{\sqrt 3} g_{818} \Tr \big( (\eta_0^\dagger \Pi_8^\dagger +
\Pi_8^\dagger
\eta_0^\dagger ) {\rm S}_8 \big),
\label{nonetham}
\end{equation}
where 
$\Pi_8=\Pi-\frac {1}{3} \Tr (\Pi)I$ and 
${\rm S}_8={\rm S} -\frac{1}{3} \Tr ({\rm S}) I $. 
The exact nonet symmetry  would require 
$g_S(K_0^*)= g_{888}=g_{818}$.  
 
For $g_{888}$ we calculate $|g_{888}|=|g_S(K_0^*)|=3.67\pm 0.3 \>{\rm GeV}$ 
from 
the partial decay width $\Gamma(K_0^*(1430)\to K\pi)$. 
In order to fix the value $g_{888}/g_{818}$  
we notice that the experimental errors might bring rather large deviation 
from the nonet symmetry. 
However, we find that the nonet symmetry  solution for $g_{888}/g_{818}=1$ 
is the average value of the allowed range and will use it from now on.

It is rather difficult to obtain the reliable estimate of the $f_{K_0^*}$. 
There are many different values in  the literature depending on the 
 model assumption. 
 The QCD sum rule estimate in the $K_0^*$ narrow width approximation 
 thus gives $f_{K_0^*} \simeq 31\pm 3\>{\rm MeV}$ \cite{Narison-87}, 
 the pole dominance result in $f_{K_0^*} \sim 50 \>{\rm MeV}$ \cite{Ayala-87}, 
 effective Lagrangian including width corrections estimate $f_{K_0^*}\sim 45 
 \>{\rm MeV}$. The extraction from the decay rate 
 $B(D^+\to \pi^+\bar{K}_0^{*0})$
using the factorization approach  results in  
$f_{K_0^*}=0.293\pm0.020$ $\>{\rm GeV}$.
Some estimates \cite{Bramon-87}, \cite{Hussain-87} 
gave $f_{K_0^*}/f_{a_0}\sim 24-30$ , while from 
$\tau \to \eta \pi \nu_\tau$ there is an upper limit $f_{a_0} < 7$ 
$\>{\rm MeV}$
\cite{Lichard-97}.
 
Having fixed all parameters  required by \eqref{art-7}, except the decay 
constant
of the $K^*_0(1430)$, we try to estimate its size 
using the  $K_{l3}$ data. 
Making low energy expansion of \eqref{art-7} we find 
\begin{align}
F_+^{21}(q^2)&=-g_V(K^*)\frac{2a_{12K^*}F_{K^*}}{m_{K^*}^2}(1+
\frac{q^2}{m_{K^*}^2} + \dots),\\
F_0^{21}(q^2)&=-g_V(K^*)\frac{2a_{12K^*}F_{K^*}}{m_{K^*}^2}(1+ 
\frac{q^2}{(m_1^2-m_2^2)} \frac{f_{K_0^*}g_S(K_0^*) c_{12K_0^*}}{F_{K^*}
g_V(K^*) 2a_{12K^*}}
\frac{m_{K^*}^2}{m_{K_0^*}^2}+\dots).
\label{art-8}
\end{align}
Comparing it with the $K_{l3}$ result in  (\ref{e1})  we obtain 
\begin{equation}
 \frac{1}{(m_1^2-m_2^2)} 
\frac{f_{K_0^*}g_S(K_0^*)}{F_{K^*}g_V(K^*)}\frac{m_{K^*}^2}{m_{K_0^*}^2}=
 \lambda_0 \frac{1}{m_\pi^2}.
\end{equation}
In the case of $K_{\mu3}^+$  $\lambda_0=0.006\pm0.007$ from which follows 
$f_{K_0^*}=0.061\pm 0.07 \>{\rm GeV}$ in fairly good agreement with 
theoretical values cited above, 
while in the case of $K_{\mu3}^0$ $\lambda_0=0.025\pm0.006$ from which 
$f_{K_0^*}=0.245\pm0.06 \>{\rm GeV}$. Due to the fairly large discrepancy between
 these numbers, we shall use $f_{K_0^*}$ as a free parameter in our further 
analysis. 

In case of $\lambda_+$ there are no such experimental ambiguities. 
The theoretical value for
\begin{equation}
\lambda_+=\frac{m_\pi^2}{m_{K^*}^2},
\end{equation}
$\lambda_+=0.0243$ is in fairly good agreement with experimental data, 
which concentrate around $\lambda_+ \sim 0.03$ \cite{PDL-98}.

We analyze final states with $S\ne 0$. Measured are 
the decay rates for 
$D^0\to \pi^0 \bar{K^0}$, $D^0\to \pi^+ K^-$, $D^0\to \eta \bar{K^0}$,
 $D^0\to \eta' \bar{K^0}$,
$D^0\to K^+\pi^-$,  while for the $D_s^+\to K^0 \pi^+$  an experimental upper 
bound exists.

We calculate invariant amplitudes using factorization approximation and form 
factors $F_{1,0}^{K\to\pi}$ in \eqref{art-8}. The invariant amplitude  is thus
\begin{equation}
\begin{split}
M_{D \rightarrow P_1P_2}
=\frac {G_f}{\sqrt{2}} & \Bigg[ 
  C_{P_1}^{D\to P_1P_2} i f_{P_2} F_0^{D \rightarrow P_1} 
(m_{P_2}^2) 
\left(m_{D}^2 - m_{P_1}^2 
\right) +\\
&C_R^{D\to P_1P_2} i f_{D}f_R \frac{m_D^2}{m_D^2-m_R^2}\langle P_1P_2|
 H_{\text{eff}}^S
| R\rangle 
\Bigg]\label{sem-25},
\end{split}
\end{equation}
where $R$ denotes the intermediate resonance, while  $C_{P_1}^{D\to P_1P_2}$ 
and $C_{R}^{D\to P_1P_2} $ are coefficients presented in 
Table \ref{tab-koef}.

\begin{table} [h]
\begin{center}
\begin{tabular}{|l|c|c|c|} \hline
$D \rightarrow P_1P_2$ & $C_{P_1}^{D\to P_1P_2}$& R  & 
$C_R^{D\to P_1P_2}$  \\  \hline\hline

$D_S^+ \rightarrow K^0 \pi^+$ & $ V_{cd}^*V_{ud}a_1$& $K_0^{*+}$&
 $V_{cs}^*V_{us}$  \\ \hline
$D^0 \rightarrow \pi^0 \bar{K}^0$ & $V_{cs}^*V_{ud}a_2/\sqrt{2}$&
$\bar{K}_0^{*0}$&$V_{cs}^*V_{ud}a_2$ \\ \hline
$D^0 \rightarrow  K^-\pi^+$ & $V_{cs}^*V_{ud}a_1$&$\bar{K}_0^{*0}$&
$V_{cs}^*V_{ud}a_2$  \\ \hline
$D^0 \rightarrow \eta \bar{K}^0  $ & 
$V_{cs}^*V_{ud}a_2(\cos(\theta_\eta)/\sqrt{6}-\sin(\theta_\eta)/\sqrt{3})$&
$\bar{K}_0^{*0}$&$V_{cs}^*V_{ud}a_2$  \\ \hline
$D^0 \rightarrow \eta' \bar{K}^0 $ & 
$V_{cs}^*V_{ud}a_2(\sin(\theta_{\eta'})/\sqrt{6}+\cos(\theta_{\eta'})/\sqrt{3})$&
$\bar{K}_0^{*0}$&$V_{cs}^*V_{ud}a_2$ \\ 
\hline
\end{tabular} 
\caption{\footnotesize{Table of coefficients in terms of  which the invariant 
amplitudes \eqref{sem-25} are expressed. The coefficients for the decays of
 $D^0$ meson conjugate to the ones listed in table are obtained by replacing 
 in $D\to \bar P_1 \bar{P_2}$ row $V_{cs}^*V_{ud}$  with $V_{cd}^*V_{us}$ and
 $R$ with $\bar R$.}}
\label{tab-koef}
\end{center}
\end{table}

In the D meson decay modes with  the neutral $K$ mesons in final state 
one has to notice that in experiment one sees 
 a short lived $K_S$ \cite{Procario-93}, 
( $K_{S,L}=1/\sqrt{2}(K^0 \mp \bar K^0)$). In the Cabbibo allowed 
decays of the type $D\to \bar K^0+X$ the annihilation contribution 
obtains the additional contribution due to the interference of the 
Cabbibo allowed $D\to \bar K^0+X$ and Cabbibo doubly suppressed 
$D\to K^0+X$ decay amplitudes that has not 
been taken into account in the analyses of experimental data. 
Thus we make a fit to the experimental data $\Gamma_{\text{PDG}}$ 
\cite{PDL-98} 
knowing that the measured decay width corresponds to $K_S$ as 
\begin{equation}
\Gamma_{\text{Exp}}(D\to K_S X)=\frac{1}{2}\Gamma_{\text{PDG}}
(D\to \bar K^0 X).
\end{equation}

For the description of the decay amplitudes in (\ref{sem-25})
we use single  pole approximation \cite{Bauer-87,Buccella-95} for the form 
factors describing  the  $D$ transitions to pseudoscalars
\begin{equation}
F_{0}^{D_{(s)}\rightarrow P}(q^2)=\frac{ F_{0}^{D\rightarrow 
P}(0)}{1-q^2/m_{D_{(s)}^*(0^+)}^2}.
\end{equation}
where for the not yet detected  scalar $D$ meson mases we take 
$m_{D^*(0^+)}=2.47$ GeV, 
$m_{D_s^*(0^+)}=2.6$ GeV \cite{Buccella-95}. The values of form factors at 
$q^2=0$ are 
$F_0^{D_S \rightarrow K} (0) =0.64 $, 
$ F_0^{D\rightarrow K} (0)=0.76$, 
$F_0^{D\rightarrow \pi}(0)=0.83$, 
$F_0^{D\rightarrow \eta}(0)=0.68$, 
$ F_0^{D\rightarrow \eta'}(0)=0.66$, 
taken from \cite{Bauer-87,Verma-95}.
 The decay constants are taken from \cite{Richman-96,Martinelli-96}
$f_D=0.21\pm 0.04 \>{\rm GeV} $, $ f_{D_S}=0.24\pm 0.04  \>{\rm GeV}$, 
$f_\pi =0.13 \>{\rm GeV}$, $f_K=0.16\>{\rm GeV}$.

The numerical results obtained for the decay widths are given in the first 
column of the Table \ref{tab:nedin}. 
We notice that our framework does not lead to 
the rates which are close enough to the observed decay rates. 
In this scheme it seems  impossible to make the  fit 
with acceptable $\chi^2$. 
Particularly, the  decay rates for  
$D^0\to K_S \pi^0$ as well as $D^0\to \eta' K_S$,$D^0\to \eta K_S$
  do not agree very well 
with the data, and therefore 
we conclude that other effects  should be taken into account.  
Following the idea of \cite{Buccella-95} we include the effects of the final 
state 
interaction.

\section{The FSI effects in $D \to K P$ decays}\label{D-meson}

It has been pointed out that in 
the D meson nonleptonic decays 
 one might expect the final state interaction to be of great importance  
\cite{Buccella-95,Close-96,Zenczykowski-97,Smith-98}. 
One way to describe FSI is through unknown phases 
attached to isospin eigenstates \cite{Bauer-87,Smith-98}. 
The phases are then determined from decay widths. The other approach was
suggested in  \cite{Buccella-95} assuming that FSI in 
two-body 
nonleptonic $D$ meson decays can be described by rescattering through scalar 
resonance $K_0^*(1950)$ \cite{Aston-88} and other 
resonances belonging to the same nonet. 
In the description of the rescattering through $K_0^*(1950)$ resonance 
we account for the $SU(3)$ symmetry breaking by taking the observed masses and 
decay widths.

The rescattering through a multiplet of resonances  mixing  different channels
 is described by \cite{Weinberg-95}
\begin{equation}
M_{N' N}^{FSI}=M_{N' N} - i \frac{\Gamma}{E-m_R+i\Gamma/2} \sum_r a_{N'}^{(r)} 
a_{N''}^{(r)*} M_{N'' N}.\label{FSI-4}
\end{equation}
Here the summation runs over various mass-degenerated resonances, 
$m_R$ is the mass of the resonance, $M_{N'N}$ is the invariant amplitude as calculated in \eqref{sem-25}, while the coefficients 
$a_{N}^{(r)}$ describe couplings of $N$-th  state to $r$-th resonance (the
summation over $N''$ is implicit). 
Coefficients $a_{N}^{(r)}$  are orthonormal
\begin{equation}
\sum_N a_N^{(r)*} a_N^{(s)}=\delta_{rs}\qquad \text{and } \quad \sum_N | 
a_N^{(r)} |^2 =1. \label{FSI-1}
\end{equation}
Since we are interested in the  final states with $S\ne0$ the only 
resonant state contributing to rescattering we account for is 
$K_0^*(1950)$. This means there is no summation over $r$ in \eqref{FSI-4}, 
while the states $N$ $(N',N'')$ are for instance in the case of 
$\bar K_0^{*0}$ : $N=\pi^0\bar K^0,
K^-\pi^+,\eta \bar K^0, \eta' \bar K^0$. From the partial decay widths  
we obtain the coefficients
\begin{equation}
a_{P_1 P_2}=\frac{ \left<P_1P_2\right| \tilde{{\cal H}}_{\text{eff}}^S \left| 
K_0^*(1950) \right> p_{P_1 P_2}^{1/2} }{\sqrt{\sum_{P_1 P_2} 
\left<P_1P_2\right|
\tilde{{\cal H}}_{\text{eff}}^S \left| K_0^*(1950) \right> ^2 p_{P_1 P_2}}}, 
\label{FSI-2}
\end{equation}
where  $p_{P_1 P_2} $ is the momentum of the final particles in the $D$ meson 
rest frame \eqref{art-27}. Here the tilde in 
$  \tilde{{\cal H}}_{\text{eff}}^S$ denotes the 
replacement of  coupling constants $g_{...}\to \tilde{g_{...}}$  in 
(\ref{nonetham}) 
as now 
pseudoscalars couple to different resonance, and the replacement $S\to 
\tilde{S}$, 
where $\tilde S$ denotes the nonet of these higher scalar resonances. 
Note  that our phases differ by a factor 2 in comparison with  $\delta_8$ in
 the equation (3.4) in \cite{Buccella-95} and are in agreement with 
\cite{Cheng-99}.
 In order to consistently treat $SU(3)$ flavor symmetry breaking 
 we keep the dependence on different masses in $p_{P_1 P_2}$, what was not 
included in 
 analyzes of 
 \cite{Buccella-95}.

Finally the decay width for various decay modes considered reads as
\begin{equation}
\Gamma_{D\rightarrow P_1 P_2}=\frac{1}{2\pi} \frac{|M_{D\rightarrow P_1 
P_2}^{FSI}|^2}{4m_D^2}p_{P_1P_2}.\label{razp-6}
\end{equation}

There are many sources of uncertainty which might arise in 
our calculation. 
For example assuming that the  nonet symmetry holds both in the case 
of $K_0^*(1430)$ as in  the case of 
$K_0^*(1950)$, we still have undetermined parameter $f_{K_0^*}$. 
The additional  uncertainties might show up due to the mixing 
(see \eqref{FSI-4})
of the four decay channels
in which resonance $K_0^*(1950)$  contributes 
(for instance $D^0\to \pi^0 \bar{K^0}$, $D^0\to \pi^+ K^-$, 
$D^0\to \eta \bar{K^0}$, $D^0\to \eta' \bar{K^0}$). If 
the amplitude for  the decay into one of the above intermediate states is 
poorly known theoretically (e.g. 
$D^0\to \eta' \bar{K^0}$ is usually considered to be 
relatively poorly known theoretically), then the uncertainty translates into 
all of the decay modes. The other problem which we encounter in the
 consideration of  the two-body  $D_{(s)}\to KX$ decays is that the 
resonance state $K_0^*(1950)$ is 
poorly known experimentally with the total decay width
$\Gamma =201 \pm 34 \pm 79\> {\rm Mev} $ and branching ratio $\Gamma(K\pi)/
 \Gamma =0.52 \pm 0.08 \pm 0.12 $.
To check how these uncertainties effect calculated decay widths we first vary
 the $K_0^*(1950)\to K \pi$ partial decay width within 
experimental bounds, while taking the whole decay width $\Gamma$ at its 
average experimental value.  This variation results in relaxing the 
requirement of the  exact nonet symmetry for the $K_0^*(1950)$ decay modes. 
We find that the  limits on nonet symmetry breaking parameter are 
$0.7<|\tilde g_{818}/\tilde g_{888}|< 1.4$. 
However, a relatively small change in $\tilde g_{818}/\tilde g_{888}$ 
results in  considerable change in decay widths.  
We found that the   best $\chi^2$ fit is realized
 using  the lower bound on $\Gamma(K\pi)/ \Gamma $.
We present  results of this fit in Table \ref{tab:nedin}.  
With the value $|g_S(K_0^*)|=3.67 \>{\rm GeV}$ calculated in previous section, 
one finds from the fit in the 
Table \ref{tab:nedin} $|f_{K_0^*}|=0.079 \>{\rm GeV}$. 
This value is close to results of  \cite{Narison-87,Ayala-87}
and close to the value obtained from $K_{\mu 3}^+$. 

We have already pointed out that 
$K_0^*(1950)$ decay parameters  have rather large errors. 
By taking into account uncertainties coming both from $\Gamma(K\pi)/ \Gamma  $ 
 and the total decay width $\Gamma$,  we extract only bounds on
$-0.13 \> {\rm GeV}<f_{K_0^*}<0.027\>{\rm GeV}$ (where $g_S(K_0^*)=3.67
 \>{\rm GeV}$ has been chosen with
positive sign). Since we cannot determine the sign of $g_S(K_0^*)$, 
we can only  limit  $|f_{K_0^*}|<0.13 \>{\rm GeV}$. 
The  bound on $|f_{K_0^*}|$ is two standard deviations away 
from the value obtained from $\lambda_0$ in $K_{\mu3}^0$ decay, 
while the value that follows from $K_{\mu3}^+$ decay is well within our 
limits (see previous section).

\begin{table} [h]
\begin{center}
\begin{tabular}{|l|c|c|c|} \hline
\raisebox{-3mm}{Decay} & $\Gamma_{\text{no FSI}} $&$\Gamma_{\text{Th}} $ & 
$\Gamma_{\text{Exp}} $  \\ 
&\scriptsize{ $(10^{-14} \>{\rm GeV}) $} &\scriptsize{$(10^{-14} \>{\rm GeV})
 $}
&  \scriptsize{$(10^{-14} \>{\rm GeV})$}  \\ \hline \hline
$D_S^+ \rightarrow K_S \pi^+  $             & 0.43& 0.17 & $ < 1.13$  \\ \hline
$D^0 \rightarrow \pi^0 {K}_S$ &  0.44&1.75 &$1.67 \pm 0.18$ \\ \hline
$D^0 \rightarrow \pi^0 {K}_L$ &  0.38&1.46 &$-$ \\ \hline
$D^0 \rightarrow  \pi^+ K^-$    & 6.29&6.41& $ 6.06\pm 0.25$ \\ \hline
$D^0 \rightarrow \eta K_S  $ &  0.13&0.29  & $ 0.55 \pm 0.09$ \\ \hline
$D^0 \rightarrow \eta K_L  $ &  0.11&0.24  & $-$ \\ \hline
$D^0 \rightarrow \eta' K_S $ &  0.19&1.22 & $ 1.35 \pm  0.22$ \\ 
\hline
$D^0 \rightarrow \eta' K_L $ &  0.13&1.02 & $ -$ \\ 
\hline
$D^0 \rightarrow K^+ \pi^- $ & 0.020&0.013 & $0.046 \pm 0.023$ \\ \hline
\end{tabular} 
\caption{\footnotesize{
The $\Gamma_{\text{no FSI}}$ denotes the calculation of 
decay widths, where we used the assumption of no final state interaction and 
fitted the parameter $f_{K_0^*}$ to $f_{K_0^*}g_S(K_0^*)=-0.46 \>{\rm GeV}^2$.
 In second 
column $\Gamma_{\text{Th}}$ are decay widths calculated in case of one 
resonance 
rescattering model of final state interaction with 
$\tilde{g}_{818}/\tilde{g}_{888}=0.7$, $f_{K_0^*}g_S(K_0^*)=-0.29
 \>{\rm GeV}^2$. 
$\Gamma_{\text{Exp}}$ denotes experimental values.
}}
\label{tab:nedin}
\end{center}
\end{table}

One can reach two  conclusions from this analysis:
first, using  the values of $f_{K_0^*}$ from the existing data
one can easily verify that the contribution of the scalar 
$K_0^*(1430)$ to the decay widths of considered $D$ meson decays is of the 
same order of magnitude as the spectator terms in factorization approximation. 
 And second, even though fairly far off-shell, the 
 contribution of the scalar projection of the 
 vector meson $K^*(892)$ is still about $20 \%$, due to the 
 large $F_{K^*}=0.195 \>{\rm GeV}^2$ decay constant and consequently the order
 of  magnitude larger $F_{K^*}g_V(K^*)$ compared to scalar
 $f_{K_0^*}g_S(K_0^*)$ (the last term 
however being enlarged by $m_D^2/(m_{P_1}^2-m_{P_2}^2)$ factor in 
\eqref{art-7}). 
 The point that annihilation terms are not negligible in decay modes 
considered is 
 best illustrated by notion that $\chi^2$ drops from $\chi^2=257$ in case of 
factorization 
 approach without annihilation terms down to $\chi^2=13$ in description with 
 annihilation terms and final state interaction included
 ($\Gamma_{\text{Th}}$ in Table \ref{tab:nedin}). As seen from the 
Table \ref{tab:nedin} also final state interaction contribute significantly 
to the decay modes considered. With the inclusion of final state interaction 
the theoretical description of decay channels in Table \ref{tab:nedin} is in a
good agreement with experimental data.

Note however that there still exists some disagreement between the 
experimental and theoretical  value of the $D^0\to \eta K_S$ decay width  for
 the best $\chi^2$ results (Table \ref{tab:nedin}). This can probably be
 attributed to the uncertainties in the description of $\eta\eta'$ system 
that show also in the  difficulties with the usual  mixing scheme 
\cite{Leutwyler-98}. Note also some discrepancy between the  average 
experimental value and our result in the doubly Cabbibo suppressed 
$D^0\to K^+\pi^-$ decay. However a sizable  uncertainty in the experimental
 result suggests that good agreement is hard to be expected.  

The uncertainties due to the experimental errors in the input parameters have 
not been included in the Table \ref{tab:nedin}, but we can roughly estimate 
that the uncertainties in $f_{D_{(s)}}$,$F_{K^*}$, $g_V(K^*)$ could result in 
the errors of about $25 \%$.

\section{$\tau \to K P \nu_\tau$ decays}\label{tau}

The same matrix element $<K P| j_{\mu}| 0>$ discussed in the previous section 
appears in the $\tau$ decay. The 
masses 
of $\tau$ and $D$ are quite close and 
the bounds on the $K^*_0(1430)$ parameters  might be tested in the $\tau$ 
decay spectra and widths. 
Unfortunately, the  data on decay spectrum are not good enough to 
obtain the information on the $K^*_0(1430)$ parameters \cite{PDL-98}. 
Here we investigate 
$\tau\to K^- \pi^0 \nu_\tau$, $\tau\to \bar{K}^0 \pi^- \nu_\tau$,
 $\tau \to K^- 
\eta \nu_\tau$ and  $\tau \to K^- \eta' \nu_\tau$
decay widths.  The first three decay widths have been measured \cite{PDL-98}, 
while  the decay $\tau \to K^- \eta' \nu_\tau$ has not yet been detected.
We start our analyses with the decay width 
\begin{equation}
\Gamma=\frac{1}{2E_1} (2\pi)^4 \delta^4 (p_1-\dots -p_n) \sum_{\rm int} 
|M_{fi}|^2 \int {\prod_{k=2}^n \frac{d^3 \vec{p}_k}{(2\pi)^3 2E_k}},
\end{equation}
where
\begin{equation}
M_{fi}=\frac{G_f}{\sqrt{2}}V_{us}J_\mu^L J_H^\mu,
\end{equation}
and $J_\mu^L=\bar{u}_{\nu_\tau} \gamma_\mu(1-\gamma_5)u_\tau$ in case of
 $\tau$ decays, while $J_\mu^H$ is hadronic current \eqref{art-4}. 
Following the procedure in \cite{Fischer-80} and denoting  $y=q^2/m_{\tau}^2$
 we obtain  the decay width
\begin{equation}
\Gamma=\int_{y_{\rm min}}^1 \frac{1}{2} \Gamma_L(1-y)^2 \big[4 
C^3(y)F_+^2(1+2y)+3C\big(\frac{m_{P_1}^2-m_{P_2}}{q^2}\big)^2
\frac{F_0^2}{y^2}\big]\,dy,
\label{art-9}
\end{equation}
where
\begin{equation}
C(q^2)=\frac{p}{\sqrt{q^2}}=\big[\frac{1}{4}-\frac{1}{2}\frac{m_{P_1}^2+
m_{P_2}^2}{q^2}+\frac{(m_{P_1}^2+m_{P_2}^2)^2}{4q^4}\big]^{1/2},
\end{equation}
with $p$ the momentum of outgoing particles in $P_1P_2$ center of mass frame, 
$\Gamma_L={G_f^2 m_{\tau}^5}/{192\pi^3}$ the pure leptonic decay rate,
 while  the  lower bound of the integral 
\eqref{art-9} equals
$y_{\rm min}=(m_{P_1}^2+m_{P_2})^2/{m_\tau^2}$.

We present our numerical results for the branching ratios in Table 3, 
comparing them with the experimental results. 
The $K^*_0(1430)$ scalar meson does not 
contribute significantly to these decay rates, as noticed recently by 
\cite{Lichard-99}. 
However, we find out that in the  
case of $\tau\to K^-\eta'\nu_\tau$ we can expect considerable enhancement of
 the scalar $K_0^*(1430)$ 
contribution to the decay width in comparison with the otherwise prevailing 
$K^*(892)$ vector 
contribution. This effect has not been taken into account in \cite{Li-97}.

\begin {figure} \begin{center}
\epsfig{file=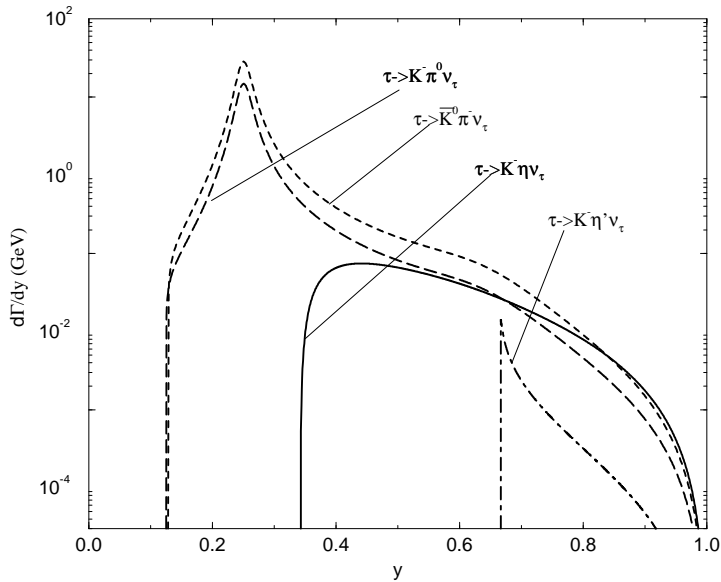,width=9cm, angle=-90}
\caption{\footnotesize{The energy dependence of decay widths for $\tau\to K^-\pi^0 
\nu_\tau$ in long-dashed, 
$\tau\to \bar{K}^0\pi^-\nu_\tau$ in dashed,
$\tau\to K^-\eta\nu_\tau$ in solid,
$\tau\to K^-\eta'\nu_\tau$ in dot-dashed line.}}
\label{sl:vselinlog}
\end{center}
\end{figure}

In the case of $\tau$ decays the final state interaction of the two outgoing 
pseudoscalar mesons can be neglected.  
The reason is that the center of mass 
energy of the pseudoscalar meson pair ranges up to the mass of $\tau$ lepton, 
with the peak of the 
distribution at the mass of $K^*(892)$, which is  below the  mass of the 
$K_0^*(1950)$ resonance. The lower bound $y_{\rm min}$ of the decay width 
energy distribution $d\Gamma/dy$ varies with the masses of the final state 
particles and is for the $\tau\to K^-\eta\nu_\tau$ with  
1.04 GeV  already above the $K^*(892)$ rest mass, 
while for the decay $\tau\to
K^-\eta'\nu_\tau$ 
the lower bound is with 1.45 GeV high above the $K^*(892)$ rest mass. 
For this decay it would be 
possible that the final state interaction would not be completely negligible, 
as the upper range of 
the decay width energy distribution is fairly near to the $K^*_0(1950)$ 
resonance. However, the 
bulk of the decay probability lies in the lower part (see 
Fig. \ref{sl:vselinlog}), thus we 
neglect the effect.
The final state interaction in $\tau\to K \pi \nu_\tau$ has been approached by 
the use of chiral perturbation theory and elastic unitarity assumption
\cite{Truong-95}, and  found to be rather small.

\begin {figure} \begin{center}
\epsfig{file=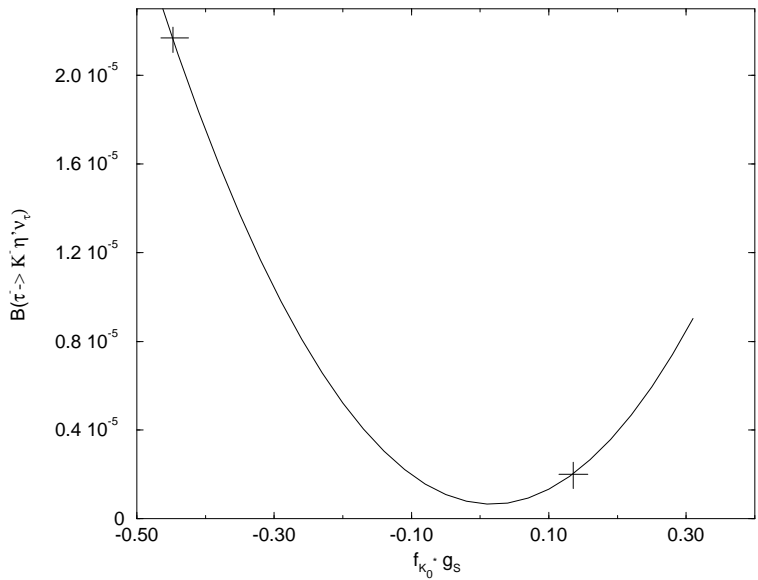,width=9cm, angle=-90}
\caption{\footnotesize{The dependence of calculated $\tau\to K^-\eta'\nu_\tau$
 decay
width  on $f_{K_0^*}g_S$, where the bounds obtained from $D\to KP$,
 $P=\pi,\eta,\eta'$ decays are denoted with crosses.}}
\label{sl:odvisnF}
\end{center}
\end{figure}

The other consequence of the different lower bounds $y_{\rm min}$ for 
different decay channels is 
that in the case of 
$\tau\to K^-\eta'\nu_\tau$ we can expect considerable enhancement 
of the scalar $K_0^*(1430)$ contribution to the decay width in comparison 
with the otherwise prevailing $K^*(892)$ vector contribution.  The size of 
this contribution crucially depends on $f_{K_0^*}$ (see Fig.\ref{sl:odvisnF}). 
This cannot be extracted inambiguosly from experimental data as discussed in 
section \ref{Decay-amplitudes}. Thus we use the value of 
$|f_{K_0^*}|=0.079 \>{\rm GeV}$ from the best fit to $D$ decays considered, to 
evaluate the size of $K_0^*(1430)$ scalar contribution, for which it is an
 order of magnitude greater than the $K^*(892)$ vector contribution 
(Table \ref{tab:tau}). The $\tau\to K^-\eta'\nu_\tau$ decay can thus be viewed 
as a possible experimental probe for the  determination of 
 $f_{K_0^*}$ decay parameter.
\begin{table} [h]
\begin{center}
\begin{tabular}{|l|c|c|c|} \hline
Decay & $B_{\text{Th}}(\%)$   & $B_{\text{Exp}}(\%) $ & scalar ($\%$)\\
\hline\hline
$\tau\to K^-\pi^0 \nu_\tau$&0.43&$ 0.52\pm 0.05$& 2.6\\\hline
$\tau\to \bar{K}^0\pi^-\nu_\tau$ &0.83 & $0.83\pm  0.08$&2.6\\\hline
$\tau\to K^-\eta\nu_\tau$ & 0.012& $0.027\pm 0.006$&6.5\\\hline
$\tau\to K^-\eta'\nu_\tau$ &0.00065&-&90 \\\hline
\end{tabular} 
\caption{\footnotesize{Branching ratios obtained for $f_{K^*}g_S(K_0^*)=-0.29 
\>{\rm GeV}^2$. The scalar denotes the size of scalar $K_0^*(1430)$ 
contribution to 
the decay width. }}
\label{tab:tau}
\end{center}
\end{table}

\section{Summary}\label{Sum}

We have reinvestigated the $D\to K P$ decays in which the annihilation 
contribution plays important role. 
We account  as much as possible for the  $SU(3)$ 
flavor  symmetry breaking effects by taking  directly from the experimental 
data all relevant parameters. Unfortunately 
 present data cannot find reliably limit on the 
$K^*_0(1430)$ decay coupling constant, therefore we kept it as a free 
parameter.
  
In the treatment of 
scalar form factor defined in the  $<KP|j_{\mu} |0> $ matrix element, 
 we included the contribution 
of the scalar projection of the vector meson $K^*(892)$ resonance. 
We found that the contribution of the 
$K^*(892)$ resonance 
in the annihilation amplitude is about $20 \%$ of the one coming from the 
scalar
meson $K^*_0(1430)$ 
resonance.

The inclusion of the final state interaction through the rescattering via 
$K^*_0(1950)$ resonance 
in these decays improves the agreement with 
the data. We then estimated bounds on the $K^*_0(1430)$ decay 
coupling $|f_{K_0^*}|<0.13\>{\rm GeV}$. 
The obtained bounds are in agreement with the experimental results obtained in 
$K_{\mu 3}^+$ decay. 
Although many other effects might be present in $D^0$ decays, like 
nonfactorizable contributions, 
we try to understand the role of scalar resonances in the annihilating 
contribution 
within this framework. 

Our analyses of the $\tau \to K P \nu_\tau$, $P= \pi, \eta$, decay spectra 
widths shows 
rather small contribution 
of the scalar $K^*_0(1430)$ resonance. 
However,  we notice its significant 
presence (about $ 90 \%$ in the rate) in the $\tau \to K \eta' \nu_\tau$ decay.
The type of the final state interaction analyzed in $D$ nonleptonic decays 
was not important in $\tau$ decays. \\
\vskip1cm

{\bf Acknowledgements}

We wish to thank B. Bajc and S. Prelov\v sek for fruitful discussions. 
This work was 
partially supported by the Ministry of Science and Technology of
 the Republic of Slovenia.

\end{document}